\documentclass[aip,amsmath,amssymb,reprint]{revtex4-1}

\usepackage{graphicx}
\usepackage{dcolumn}
\usepackage{bm}
\usepackage{ulem}
\usepackage[utf8]{inputenc}
\usepackage[T1]{fontenc}
\usepackage{mathptmx}
\usepackage{etoolbox}
\usepackage{xcolor}
\usepackage{multirow}
\makeatletter
\def\@email#1#2{%
 \endgroup
 \patchcmd{\titleblock@produce}
  {\frontmatter@RRAPformat}
  {\frontmatter@RRAPformat{\produce@RRAP{*#1\href{mailto:#2}{#2}}}\frontmatter@RRAPformat}
  {}{}
}%
\makeatother
\begin{document}

\preprint{}

\title[Characteristics of price related fluctuations in Non-Fungible Token (NFT) market]{Characteristics of price related fluctuations in Non-Fungible Token (NFT) market}

\author{Pawe{\l} Szyd{\l}o}
\author{Marcin W{\k a}torek}
 \email{marcin.watorek@pk.edu.pl}

\affiliation{Faculty of Computer Science and Telecommunications, Cracow University of Technology, ul.~Warszawska 24, 31-155 Krak\'ow, Poland}

\author{Jaros{\l}aw Kwapie\'n}

\affiliation{Complex Systems Theory Department, Institute of Nuclear Physics, Polish Academy of Sciences, ul.~Radzikowskiego 152, 31-342 Krak\'ow, Poland}

\author{Stanis{\l}aw Dro\.zd\.z}

\affiliation{Faculty of Computer Science and Telecommunications, Cracow University of Technology, ul.~Warszawska 24, 31-155 Krak\'ow, Poland}
\affiliation{Complex Systems Theory Department, Institute of Nuclear Physics, Polish Academy of Sciences, ul.~Radzikowskiego 152, 31-342 Krak\'ow, Poland}

\date{\today}

\begin{abstract}
A non-fungible token (NFT) market is a new trading invention based on the blockchain technology which parallels the cryptocurrency market. In the present work we study capitalization, floor price, the number of transactions, the inter-transaction times, and the transaction volume value of a few selected popular token collections. The results show that the fluctuations of all these quantities are characterized by heavy-tailed probability distribution functions, in most cases well described by the stretched exponentials, with a trace of power-law scaling at times, long-range memory, persistence, and in several cases even the fractal organization of fluctuations, mostly restricted to the larger fluctuations, however. We conclude that the NFT market - even though young and governed by a somewhat different mechanisms of trading - shares several statistical properties with the regular financial markets. However, some differences are visible in the specific quantitative indicators. 

\end{abstract}

\maketitle
\begin{quotation}
Over the last dozen years researchers have had a unique opportunity to observe the emergence and evolution of new financial markets. The cryptocurrency market was created in 2011 and passed through a few stages from infancy, through a stage of being an emergent market with not fully developed features, to a relatively mature stage at present. It has already been a subject of a number of studies which revealed much information about its structure and dynamics. This is not the case of the even younger and less developed non-fungible token (NFT) market, which was established only in 2017. Its short history and low liquidity make the amount of data available not yet spectacularly rich but already sufficient to make rather reliable estimates towards establishing the related analogs of the so-called stylized facts. The present work is among the first that focus on the statistical properties of the NFT market trading.
\end{quotation} 

\section{Introduction}

Introduction of a blockchain technology proposed in a seminal paper by Satoshi Nakamoto~\cite{NakamotoS-2009a} soon gave birth to a completely new financial market on which cryptocurrencies were traded~\cite{watorek2021}. Since its infancy stage in the years 2011-2015, where both prices and the number of traders were relatively small~\cite{Zunino2018} this market undergo significant evolution and now it can be considered a large, mature market that couples itself to or decouples from the regular financial markets~\cite{WatorekM-2023a,JAMES2022PhysD}. Interest in the cryptocurrency market grew enormously during the Covid-19 pandemic, because investors were looking for a safe haven in times of market panic~\cite{James2021chaos,Arouxet2022,Almeida2023,Garcin2023,JAMES2023PhysA}, low interest rates, and quantitative easing~\cite{SarkodieSA-2022a}, and this contributed substantially to the cryptocurrency market liquidity~\cite{CorbetS-2022a,James2022,Nguyen2022,James2023,Nguyen2023}. However, even though such expectations were based on actual observations that during 2019 the cryptocurrency market was largely independent from other markets~\cite{DrozdzS-2019a}, they proved futile as the cryptocurrencies had tendency to align with the main financial instruments during the most dramatic market events in 2020~\cite{DrozdzS-2020a,Manavi2020,James2021,James2021b,Almeida2023b,Karagiannopoulou2023} and recently in 2022 due to inflation concerns~\cite{James2022inf,WatorekM-2023a,Watorek2023chaos}.

Cryptocurrencies share many traits with fiat currencies, among which one can point out to the full equivalence (indistinguishability) of units. This is what discerns cryptocurrencies from non-fungible tokens (NFTs)~\cite{GuidiB-2023a}. A non-fungible token is a type of data stored on a blockchain that contains its unique sign. In contrast to typical cryptocurrency units like bitcoin or ether, which are indistinguishable, interchangeable, and have the same value, each NFT has its individual value and cannot be replaced by another token. Ownership and authenticity of a particular token are permanently recorded and easily verifiable via a relevant blockchain so it is impossible to counterfeit it. This property can be exploited to represent ownership of digital or physical assets, such as art, music, videos, games, collectibles, and much more~\cite{RegnerF-2019a,CaglayanAksoyP-2021a,TaherdoostH-2023a}.

The idea of NFT was put into practice with the adoption of ERC-721 smart-contract protocol that allowed users of Ethereum blockchain to create tokens that could have distinguishable signs~\cite{ERC-721,IbanezL-2018a}. While originally confined to this network only, later on it spread to other blockchains like Solana, Polygon, Flow, Binance Smart Chain, Cardano, Arbitrum, and recently even Bitcoin~\cite{bertucci2023bitcoin}. Although the first NFT was presented in May 2014, a significant trigger for the NFT market development was the launch of the CryptoKitties collection in November 2017. CryptoKitties is an online game in which a player can buy, breed, and sell virtual cats represented by cartoon images. Each ``cat'' is a non-fungible token that differs from other tokens by its visual properties and rarity. The game operates on the Ethereum blockchain network and at one moment it became so popular that it caused an overload of the whole network~\cite{Jiang2021,Serada2021}. Shortly thereafter, the first organized NFT marketplace - OpenSea - began operating. The NFT-market has boomed in recent years even more, especially during the middle stage of the Covid-19 pandemic in 2021, with its peak of interest at the turn of 2021 and 2022. In October 2023, while this article was written, daily turnover of the NFT markets oscillated around 30 billion USD, which is a serious downfall if compared with the period of the record-high activity in late 2021, when turnover used to exceed 200 billion USD~\cite{CoinGecko}. There is evidence that the growth of the NFT market has closely been related to the recent bubble on the cryptocurrency market~\cite{PintoGutierrezC-2022a,Guo2023}. On the other hand, individual collection price dynamics were often shaped by the so-called NFT influencers~\cite{HORKY2023}. This was particularly visible in the highest volume traded NFT collection - Bored Ape Yacht Club, whose price changes were influenced by top celebrities' tweets~\cite{vanSlootn2022}.

An important part of the NFT market is devoted to the gaming community for selling and buying in-game items, but there are also other notable NFT types, which comprise virtual landscapes, virtual real estate, and digital artwork~\cite{giannoni2023blockchain}. It often happens that tokens and token collections are created and originally sold in order to raise funds for financing new projects. After an initial offer, NFTs can be freely traded, which poses an incentive to consider them as normal investment instruments. Indeed, many token collections allowed their primary buyers to earn a multiple of the invested amounts. The most expensive NFT was Mike Winkelmann's ``Everydays: The First 5000 Days'' -- a collection of 5000 unique illustrations that the artist had created over 5000 consecutive days, which was sold for 69 million USD in March 2021~\cite{Christies}. This event drew attention to the growing popularity of the NFT market and the significance of digital assets in today's art world. After a few years of existence of the NFT technology, Ethereum remains the blockchain with the largest market share and it is followed by others: Bitcoin, ImmutableX, Polygon, Solana, Flow, Arbitrum and BNB Chain with a changing share in the market capitalization and trading volume~\cite{CoinGecko}.

The NFT market is second to the cryptocurrency market in terms of research interest to date~\cite{BaoH-2022a,GuidiB-2023a}. It is because novelty of NFTs and their non-fungible character bear several challenges for NFT market research~\cite{ChoJB-2023a,AliO-2023a}. The first challenge is that the NFT market is younger and substantially less liquid than the cryptocurrency market. It also reveals extreme volatility that together with infrequent trading seriously limits the data volume that can be analyzed~\cite{ChoJB-2023a}. The second challenge is high heterogeneity of token prices within a collection. It is partially associated with a token's rarity, i.e., a relative frequency of occurrence of the traits attributable to this token. It has been reported that token rarity may have an impact not only on its price (rare tokens are expensive) but also on a frequency of trading (rare tokens are seldom traded), amplitude of return (rare tokens are profitable), and negative-return resilience (rare tokens are safer)~\cite{MekacherA-2022a}. Although these properties of tokens make statistical analysis difficult, some of them may however constitute good predictive factors regarding future price~\cite{NadiniM-2021a}. The third challenge are lateral swaps, which can distort price evolution of a given token by substituting, for example, a smart contract execution fee that differs substantially from a token's market price that should actually be listed there. The fourth challenge is wash trading, i.e., a long series of trading of the same token, a goal of which is to deceive other market participants about actual liquidity and/or price of this token. Wash trades are extremely difficult to be filtered out, because they may not differ from the proper trades in terms of price and frequency. It was suggested that a majority of token collections undergo this procedure from time to time~\cite{vonWachterV-2022a}.

High volatility and the related inefficiency of the NFT market was the main outcome of a study of pricing of LAND tokens representing parcels of virtual estate in a metaverse of Decentraland~\cite{DowlingM-2022a}. This inefficiency manifested itself in dominating antipersistence of daily LAND price which can potentially allow one to make related profits. However, such a behaviour may always be expected in the case of a young market. High volatility of NFT prices does not exclude using these assets for hedging, especially against the more risky cryptocurrencies and DeFis~\cite{KarimS-2022a}. It was shown that net volatility spillover in low- and medium-volatility events is largely positive for cryptocurrencies and DeFis, while it is negative for NFTs~\cite{KarimS-2022a}. As it is a measure of risk transmission potential, NFTs have a potential for diversifying portfolio. This is not the case, however, if a high-volatility events are considered. A conclusion that goes in parallel direction was given in~\cite{AharonDY-2022a} according to which NFTs generate their own shocks and are rather resistant to external ones; this refers even to ether, which is closely related to many NFTs. During turbulent times NFTs tend to absorb external volatility shocks, while during times of normal evolution they behave like transmitters of risk spillovers~\cite{AharonDY-2022a}.

Somehow opposite conclusion can be learnt from a study of the cryptocurrency, DeFi, and NFT bubbles that occurred during the pandemic~\cite{MaouchiY-2022a}. According to that study, the NFT and DeFi bubbles grow faster and reach higher levels, but also they are less frequent than cryptocurrency bubbles. Also outside the bubbles NFTs show high-return and high-volatility profile that make them ideal for the investors accepting risk in order to make quick gains~\cite{KongDR-2021a}. An increased coherence between evolution of NFTs and regular financial assets like bonds, equities, and commodities (gold, oil) was reported~\cite{UmarZ-2022a,AharonDY-2022a} during the pandemic (2020), which agrees with a parallel conclusion regarding the cryptocurrency market~\cite{DrozdzS-2020a}. A different study suggests that NFTs reveal certain degree of coherence with cryptocurrencies, but it must be limited as volatility transmission between the two is weak~\cite{DowlingM-2022b}.

Despite the above research works, there is a limited number of quantitative studies focusing on the statistical properties of NFTs. In this article we want to fill this gap partially through an analysis of a few selected financial observables like inter-transaction times, return distributions, and transaction volume values~\cite{Watorek-2021entropy,watorek2021,KwapienJ-2022a,WatorekM-2023a,WatorekM-2023b}. We focus on a few sample NFT profile-picture collections that are distinguished by their popularity among the NFT investors that was manifested by a high transaction volume around mid-2023. Below there are short descriptions of all collections considered in this study. Each of these collections is operated on the Solana blockchain network.

\textit{Blocksmith Labs Smyths} collection is a set of 4,444 NFTs associated with \$FORGE token that have been created and operated on Solana blockchain network. They are represented by digital images depicting certain noble characters with fantastic attributes. The collection was started in March 2022 by the company Blocksmith Labs, which specializes in development of products exploiting the blockchain technology~\cite{BlocksmithLabs}. \textit{Famous Fox Federation} collection is a set of 7,777 NFTs associated with FOXY token. They are represented by digital images of stylized cartoon foxes with different characteristics. The collection was developed by a group of Solana network users in order to help them create an ecosystem of various blockchain-related tools~\cite{FamousFoxFederation}. \textit{Lifinity Flares} collection consists of 10,000 NFTs represented by animated images of flares. It was released in December 2021 by Lifinity Foundation in order to fund development of its market making protocol aimed at improving liquidity and increasing investor revenues~\cite{Lifinity}. \textit{Okay Bears} is a collection consisting of 10,000 NFTs represented by digital images of cartoon bears with different colour and outfits. The collection, which was launched by two private entrepreneurs, is associated with a broadly planned project whose goal is to be a virtual place for hosting various community and start-up initiatives~\cite{OkayBears}. Finally, \textit{Solana Monkey Business} collection is a set of several thousand NFTs that are represented graphically by pixel-art images of monkeys and grouped in 3 ``generations''. These NFTs have been being released in tranches since May 2021 by a MonkeDAO community specializing in development of Web3 projects~\cite{MonkeDAO}.

\section{Data and methods}
\label{sect::data.and.methods}

We consider tick-by-tick data representing transactions (price and time) on 5 NFT collections: Blocksmith Labs Smyths (BL), Famous Fox Federation (FF), Lifinity Flares (LF), Okay Bears (OKB), and Solana Monkey Business (SM) and floor prices of the same collections. Each data set covers a period starting on a collection launch day and ending on Sep 1, 2023. We transformed these data sets into time series of 5 observables: collection capitalization $C$, floor price $p_{\rm fl}$, transaction value $V_{\Delta t}$ (volume multiplied by price) in time unit $\Delta t$, the number of transactions in time unit $N_{\Delta t}$ and inter-transaction times $\delta_{t_i}=t_{i+1}-t_i$. Because over a few days after the market debut by a collection trading is incomparably more intense than later on, in order for our analysis to be feasible, we omit the data representing the first week of trading for each collection. Without this step the data would have been too non-stationary. A relatively small number of transactions that are made on NFT collections and long waiting times between the consecutive transactions as well as value variability among the members of the same collection has led to the need for a different measure of collection value than a last transaction price. Floor price is defined for a collection and it refers to the lowest ask price of a token belonging to this collection that sellers can accept at a given moment. Floor price may differ from the lowest transaction price and it is a unique characteristic of the NFT market. This feature makes the NFT market more closely resemble an auction market than traditional financial markets with a book of buy and sell orders.

The data has been downloaded from CryptoSlam! portal (tick-by-tick transaction data)~\cite{CryptoSlam} and Magic Eden portal (floor price data), which is an aggregator of the most important Solana marketplaces: CoralCube, Elixir, Fractal, HadeSwap, OpenSpace, Solanart, TensorSwap~\cite{MagicEden}. Selected characteristics of each collection are gathered in Tab.~\ref{tab::collection.stats}. Based on transaction prices, for each collection, we calculated total collection capitalization by summing up the last transaction prices of all the tokens expressed in USD and SOL. In order to make the data more stationary, we transformed the tick-by-tick data of collection capitalization and floor price into time series of logarithmic capitalization increments $c_{\Delta t}(t)=\ln C(t+\Delta t) -\ln C(t) $ and floor price returns $r_{\Delta t}(t)=\ln p_{fl}(t+\Delta t) - \ln p_{fl}(t)$ with sampling frequency $\Delta t$, which we chose to be 1 hour, because the transaction frequency on the NFT market is rather low. As value of solana, the native cryptocurrency for the considered collections, fluctuates with respect to the US dollar, four time series were subject to further analysis: collection capitalization increments for USD and SOL and collection floor price returns for USD and SOL.

In order to inspect the self-similarity properties of the time series, we applied the multifractal detrended fluctuation analysis (MFDFA), which is a relatively simple and reliable method to detect scaling of the fluctuations~\cite{KANTELHARDT2002,OswiecimkaP-2006a}. For time series $u(j)$ let ${\rm X} = \{X(i)\}_{i=1}^T$ be an integrated time series:
\begin{equation}
X(i) = \sum_{j=1}^i u(j).
\end{equation}
For a given scale $s$, we divide X into $M_s$ disjoint segments of length $s$ starting from both ends, which gives $M_s=2\lfloor T/s \rfloor$, and fit the $m$th-degree polynomial $P^{(m)}_{s,\nu}$ to X in each segment $\nu$. Then we subtract a local trend defined by this polynomial from the signal profile:
\begin{equation}
x_{s,\nu}(i) = X_{s,\nu}(i) - P^{(m)}_{s,\nu}(i)
\end{equation}
and calculate variance of $x$ in each segment separately:
\begin{equation}
f^2(s,\nu) = {1 \over s} \sum_{i=1}^s (x_{s,\nu}(i) - \langle x_{s,\nu} \rangle)^2.
\end{equation}
In the next step, variance $f^2$ is averaged over all segments and the $q$th fluctuation function is calculated:
\begin{equation}
F_q(s) = \left\{ {1 \over M_s} \sum_{\nu=0}^{M_s-1} \left[ f^2(s,\nu) \right]^{q/2} \right\}^{1/q},
\label{eq::fluctuation.function}
\end{equation}
where $q \in \mathcal{R}$. By repeating the above procedure for a number of different scales and values of $q$, one can observe how the function $F_q(s)$ behaves for different $s$. It occurs that for fractal signals $u(j)$ the fluctuation function shows a power-law form for all $q$s and for a broad range of scales $s$:
\begin{equation}
F_q(s) \sim s^{h(q)},
\end{equation}
where $h(q)$ depends monotonically on $q$ and it is called a generalized Hurst exponent. For a constant $h(q)=H$, the signal $u(j)$ is monofractal, otherwise it is multifractal. Since its introduction, MFDFA has been widely adopted in financial data analysis. Its application to empirical data from various markets made it possible to show that price returns and inter-transaction waiting times reveal multifractal structure~\cite{FisherA-1997a,CalvetL-2002a,OswiecimkaP-2005a,KwapienJ-2005b,CajueiroD-2009a,RuanY-2011a,RAK2018,jiang2019multifractal,Klamut2020}. Recently, it has been shown that the same property can be identified in cryptocurrency market data~\cite{DrozdzS-2018a,takaishi2018statistical,DrozdzS-2019a,kristj2019,han2020long,takaishi2020market,BARIVIERA2021,Takaishi2021,watorek2021,KwapienJ-2022a,KakinakaS-2022a,Watorekfutnet2022}.

\begin{table}
\caption{Basic characteristics of the NFT collections considered in this study: the number of circulated tokens $K$, time series length $T$, the average inter-transaction time $\langle \delta t \rangle$, the average transaction number $\langle N_{\Delta t} \rangle$, the average transaction volume value in SOL $\langle V_{\Delta t}^{\rm SOL} \rangle$, and the fraction of zero log-returns for total collection capitalization $\%0 c_{\Delta t}$ and for floor price $\%0 r_{\Delta t}$, for $\Delta t=1$ h.}
\label{tab::collection.stats}
\scriptsize
\begin{tabular}{|l|c|c|c|c|c|c|c|c|}
\hline
Collection& Start date & \textit{K} & \textit{T} & $\langle \delta t \rangle$ [s] & $\langle N_{\Delta t} \rangle$ & $\langle V_{\Delta t}^{\rm SOL} \rangle$ & \textit{$\%0 c_{\Delta t}$} & \textit{$\%0 r_{\Delta t}$} \\ \hline\hline
BL   & 03/24/2022 & 4,444 & 12,627 & 1,888 & 1.23 & 51.60 & 0.47 & 0.56 \\ \hline
FF    & 09/30/2021 & 7,051 & 16,826 & 1,517 & 1.85 & 42.94 & 0.32 & 0.49 \\ \hline
LF          & 12/25/2021 & 9,963 & 14,763 & 1,926 & 0.89 & 4.03 & 0.71 & 0.81 \\ \hline
OKB               & 04/26/2022 & 9,925 & 11,834 & 777  & 2.78 & 151.10 & 0.29 & 0.51 \\ \hline
SM   & 08/16/2021 & 3,910 & 17,902 & 4,102 & 0.82 & 48.63 & 0.62 & 0.80 \\ \hline 
\end{tabular}
\end{table}

\begin{figure*}
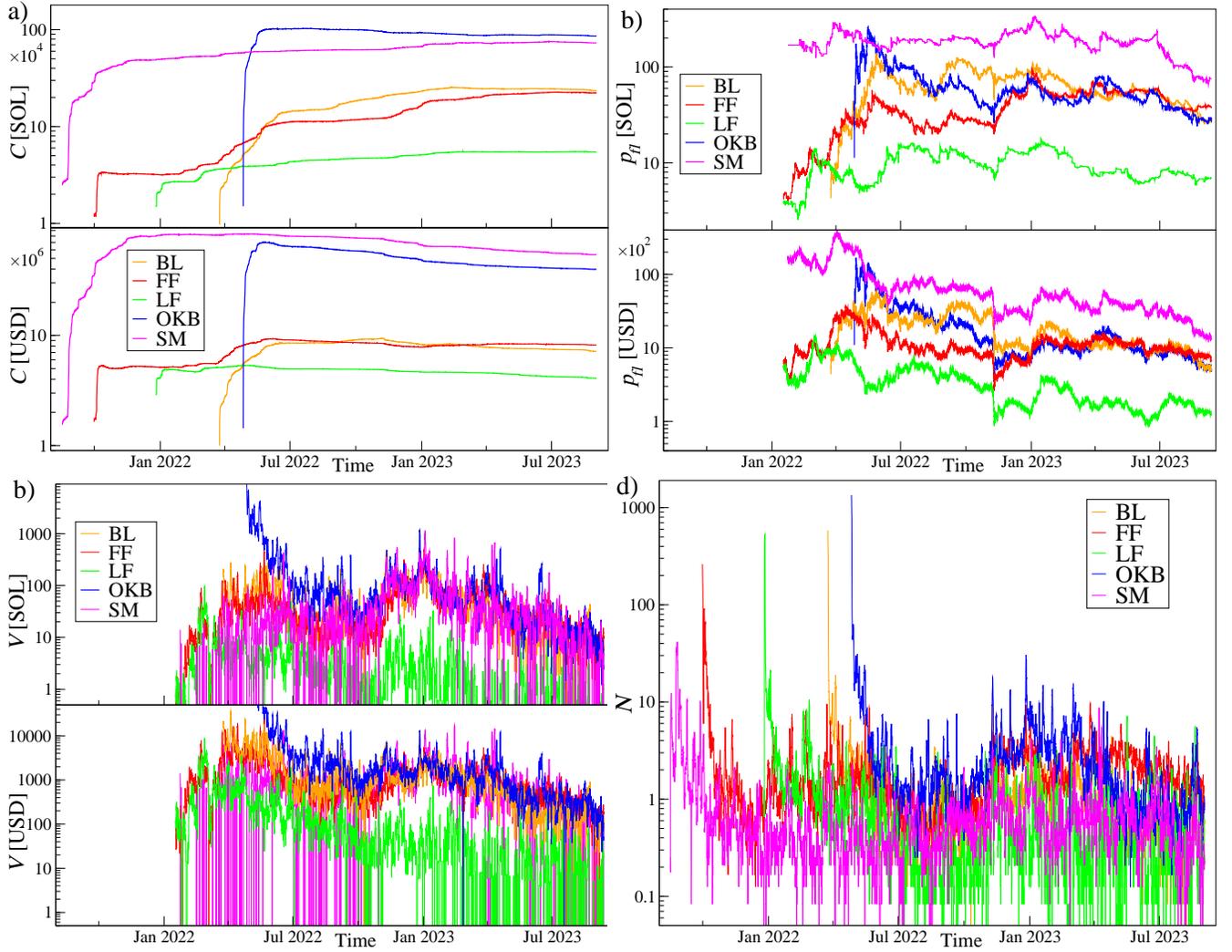

\includegraphics[width=0.49\textwidth]{figs/stat_kolekcjiallC.eps}
\includegraphics[width=0.49\textwidth]{figs/stat_kolekcjialluRfl.eps}

\includegraphics[width=0.49\textwidth]{figs/stat_kolekcjialluV.eps}
\includegraphics[width=0.49\textwidth]{figs/stat_kolekcjialluN.eps}
\caption{Collection characteristics: (a) total capitalization $C$ expressed in SOL and USD, (b) transaction volume value $V_{\Delta t}$ in SOL and in USD, (c) floor price $p_{\rm fl}$ in SOL and in USD, (d) transaction number aggregated hourly $N_{\Delta t}$ for all 5 collections: Blocksmith Labs Smthys (BM), Famous Fox Federation (FF), Lifinity Flares (LF), Okay Bears (OKB), and  Solana Monkey Business (SM).}

\label{fig::stat}
\end{figure*}

Let us begin a presentation of the results by displaying the time series themselves. Fig.~\ref{fig::stat} show time series representing the analysed observables for each of the 5 NFT collections studied. These are: (a) the total market capitalization $C$ of a collection, (b) floor price $p_{\rm fl}$, (c) the 1-hour transaction volume value $V_{\Delta t}$,  and (d) the aggregated hourly number of trades $N_{\Delta t}$ (please not that, to some approximation, $N_{\Delta t}$ and $\delta t$ are variables that are inverse of each other). Given that SOL, like other cryptocurrencies, is one of the speculative assets with high price fluctuations, it is useful to express the capitalization of the collection in both SOL and USD ($C_{_{\rm SOL}}$, $C_{_{\rm USD}}$, respectively - Fig.~\ref{fig::stat}(a)). It can clearly be seen that behaviour of the two quantities is fundamentally different. While in the first few months after the debut both $C_{_{\rm SOL}}$ and $C_{_{\rm USD}}$ increase rapidly over time, which is related to the novelty effect and the token-release-related news, after this initial period the curves representing the two quantities show quite different behaviour. In the case of $C_{_{\rm SOL}}$, the increase in value of 4 out of 5 collections continues even for about a year after the debut, while for $C_{_{\rm USD}}$ there was a saturation of the capitalization and then a slow decline. This, of course, was related with the behaviour of the two instruments' mutual exchange rate: in 2022 there was a downturn in the cryptocurrency market and the value of SOL expressed in USD used to decrease gradually. A different picture emerges in the case of the OK Bears collection, however, where both $C_{_{\rm SOL}}$ and $C_{_{\rm USD}}$ started to trend downwards after a few initial months.

The strong interest in collections just launched is reflected in the number of transactions made per unit time $N_{\Delta t}$. For each collection analysed here, this number was many times higher immediately after a collection debut for one or more days than in later periods (Fig.~\ref{fig::stat}(d)). After these few days of trading frenzy, traffic on a given collection calms down and then it fluctuates as does the interest in the NFT market as a whole. The number of transactions varies in time, however, with significant amplification of trading frequency at the end of 2021 and Q1 of 2022.

Fig.~\ref{fig::stat}(b) shows the evolution of floor price expressed in SOL and USD. In the former case, $p_{\rm fl}$ is more stable in the sense that there is no sharp decline after an initial period of a collection-release-related euphoria. In the latter case, on the other hand, floor price quickly falls to a value several times lower than its maximum. This resembles the behaviour of the total capitalization of individual collections. However, even in the case of floor price expressed in SOL, a strong oscillation around a long-term trend is visible, which is more often horizontal and (less often) rising. Fig.~\ref{fig::stat}(c) shows the evolution of hourly transaction volume values in SOL and USD. $V_{\Delta t}$ experienced four phases in the period under study: in the first half of 2022 and in the period from the end of 2022 to the end of Q1 of 2023, trading activity was relatively high, while in the period from mid 2022 to autumn 2022 and from spring 2023 onwards, the activity decreased markedly. At a glance, one can already see the fluctuation-clustering phenomenon, which resembles a similar effect observed in the other financial markets. We therefore expect that this observable shows a long memory.

\section{Results}

We examined the empirical time series presented above for several properties that are observed typically in the other financial markets: fluctuation distributions, autocorrelation functions, and scaling properties.

\subsection{Statistical properties of distributions}

\begin{figure}
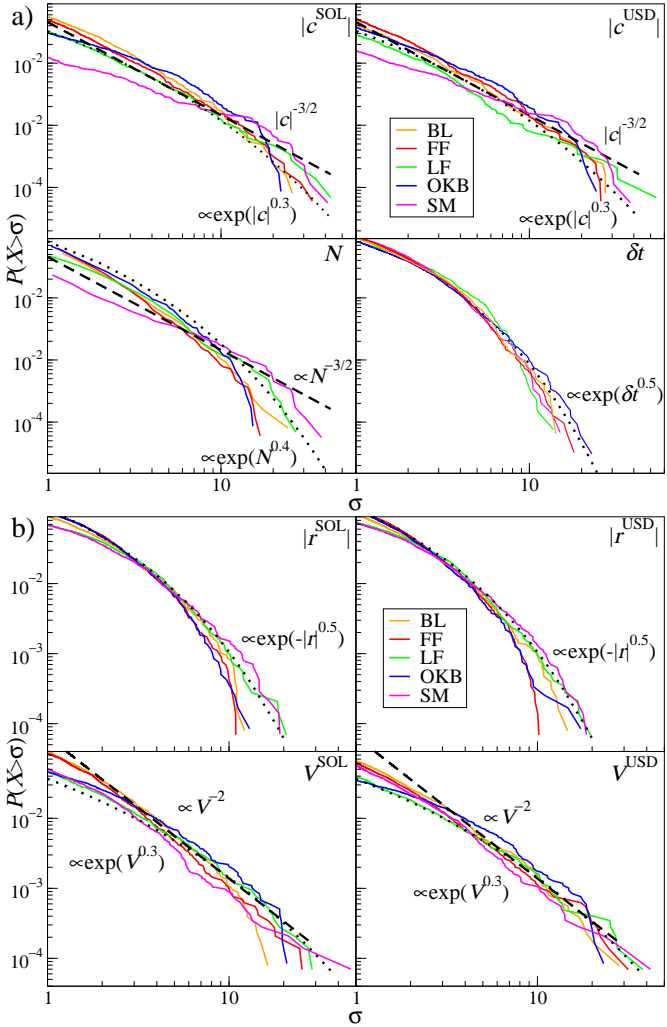

\includegraphics[width=0.49\textwidth]{figs/Rozklady_kolekcje.eps}

\includegraphics[width=0.49\textwidth]{figs/Rozklady_floorprice.eps}
\caption{(a) Probability distribution functions for absolute values of logarithmic increments of collection capitalization expressed in solana $|c_{\Delta t}^{^{\rm SOL}}|$ (top left) and US dollar $|c_{\Delta t}^{^{\rm USD}}|$ (top right), the number of transactions aggregated hourly $N_{\Delta t}$ (bottom left), and inter-transaction times $\delta t$ (bottom right). (b) Probability distribution functions for absolute values of floor price returns expressed in SOL $|r_{\Delta t}^{^{\rm SOL}}|$ (top left) and USD $|r_{\Delta t}^{^{\rm USD}}|$ (top right) and volume value expressed in solana $V_{\Delta t}^{^{\rm SOL}}$ (bottom left) and US dollar $V_{\Delta t}^{^{\rm USD}}$ (bottom right). The appropriate power law (long-dashed) and stretched-exponential (short-dashed) models are also shown as guides for an eye together with their parameter values $\gamma$ and $\beta$, respectively.}
\label{fig::distributions}
\end{figure}

Fig.~\ref{fig::distributions} shows cumulative probability distribution functions $P(X>\sigma)$, where $\sigma$ denotes standard deviation, for 8 types of time series: logarithmic increments of total capitalization $|c^{^{\rm SOL}}_{\Delta t}|$ and $|c^{^{\rm USD}}_{\Delta t}|$ (top left and top right panels in Fig.~\ref{fig::distributions}(a), respectively), the number of trades per unit time $N_{\Delta t}$ with $\Delta t=1$ h (bottom left panels), the inter-transaction times $\delta t$ (bottom right), floor price returns $|r^{^{\rm SOL}}_{\Delta t}|$ and $|r^{^{\rm USD}}_{\Delta t}|$ (top left and top right panels in Fig.~\ref{fig::distributions}(b), respectively) and volume value per unit time $V_{\Delta t}^{\rm SOL}$ and $V_{\Delta t}^{\rm USD}$ (bottom left and bottom right panels in Fig.~\ref{fig::distributions}(b), respectively). In each panel, the most plausible model curves for the data sets considered are shown as reference distributions for the respective empirical distributions. These are either distributions with power tails $x^{-\gamma}$ represented by straight lines (the power laws take a linear form on a double logarithmic scale) or stretched-exponential distributions described by a function $f(x) \approx {\rm exp} (x^{\beta})$, where $0 < \beta < 1$~\cite{LaherrereJ-1998a,MalevergneY-2005a}. These two models are the most widely applied in the context of financial data, therefore we shall restrict the analysis to these models, while different models can also be considered in general yet they can less likely represent the data. Depending on a given quantity, the model distributions have different values for the power-law exponent $\gamma$ and the stretched-exponential function exponent $\beta$. The power-law model can be applied to the time series of $|c_{\Delta t}^{\rm SOL}|$ for the FF and LF collections and $|c_{\Delta t}^{\rm USD}|$ for the BL, FF, and LF collections. The power-law exponents in all these cases are close to $\gamma=3/2$ (see Tab.~\ref{tab::model.parameters}) and this has been marked in Fig.~\ref{fig::distributions}(a) by long-dashed straight lines. In a similar way, the power-law model with exponent $\gamma \approx 2$ can be used to describe the p.d.f.s of the collection volume value $V_{\Delta t}^{\rm SOL}$ for FF and SM as well as $V_{\Delta t}^{\rm USD}$ for BL, FF, and SM as it is seen in Fig.~\ref{fig::distributions}(b) (long-dashed lines) and Tab.~\ref{tab::model.parameters}. These are the only time series for which a power-law behaviour of the empirical distribution tails can be observed; in all other cases, no scaling region can be identified.

As far as the stretched exponential functions are concerned, more empirical distributions can be well approximated with them than in the case of the power-law functions. More specifically, the p.d.f.s of $|c_{\Delta t}^{\rm SOL}|$ and $|c_{\Delta t}^{\rm USD}|$ for BL, FF, and OKB as well as the p.d.f.s of $V_{\Delta t}^{\rm SOL}$ for LF and OKB and $V_{\Delta t}^{\rm USD}$ for BL, LF, and OKB are the best described by the exponents $\beta \approx 0.3$, the p.d.f.s of $N_{\Delta t}$ for BL, FF, LF, and OKB and $V_{\Delta t}^{\rm SOL}$ for BL can be approximated with $\beta \approx 0.4$, the p.d.f.s of $\delta t$, $|r_{\Delta t}^{\rm SOL}|$, and $|r_{\Delta t}^{\rm USD}|$ for all the collections can be approximated with $\beta \approx 0.5$. A distinguished exception from this pattern, where time series of the same observable for different collections reveal comparable values of $\beta$, is seen for $N_{\Delta t}$, where $\beta \approx 0.25$ for SM while $\beta \approx 0.4$ for the other collections. Typical models for each observable are denoted by short-dashed lines in Fig.~\ref{fig::distributions}. For a few time series both models give fits of comparable quality, which means that we cannot chose a single optimal model in these cases. It is worth stressing that in neither case we can speak of a normal distribution since the tails of the empirical distributions are leptokurtic.

\begin{table}[]
\caption{Values of the power-law exponent $\gamma$ and stretched-exponential function parameter $\beta$ least-square-fitted to empirical time series together with their standard errors. Empty cells correspond to the situation where it is impossible to fit either model.}
\scriptsize
\label{tab::model.parameters}
\begin{tabular}{|l|l|c|c|c|c|c|}
\hline
\multicolumn{2}{|c|}{} & BL & FF & LF & OKB & SM \\ \hline
\multirow{2}{*}{\textbf{$|c_{\Delta t}^{\rm SOL}|$}} & $\beta =$          & $0.34        \pm 0.02$ & $0.33 \pm 0.04$        &             & $0.33 \pm 0.04$         &             \\ \cline{2-7} 
                                      & $\gamma =$         &             & $1.85 \pm 0.14$       & $1.59 \pm 0.14$       &              &             \\ \hline
\multirow{2}{*}{\textbf{$|c_{\Delta t}^{\rm USD}|$}} & $\beta =$           & $0.30 \pm 0.07$        & $0.33 \pm 0.07$        &             & $0.29 \pm 0.07$         &         \\ \cline{2-7} 
                                      & $\gamma =$        & $1.55 \pm 0.3$       & $1.64 \pm 0.3$       & $1.55 \pm 0.38$  &            &     \\ \hline
\textbf{$|r_{\Delta t}^{\rm SOL}|$}               & $\beta =$            & $0.5 \pm 0.03 $       & $0.56 \pm 0.02$        & $0.49 \pm 0.04$        & $0.55 \pm 0.02$         & $0.46 \pm 0.05$        \\ \hline
\textbf{$|r_{\Delta t}^{\rm USD}|$}                & $\beta =$          & $0.51 \pm 0.02$        & $0.55 \pm 0.04$        & $0.48 \pm 0.07$        & $0.56 \pm 0.04$         & $0.46 \pm 0.03$        \\ \hline
\textbf{$N_{\Delta t}$}                            & $\beta =$            & $0.39 \pm 0.04$        & $0.42 \pm 0.02$        & $0.36 \pm 0.05$        & $0.41 \pm 0.02$         & $0.25 \pm 0.08$        \\ \hline
\textbf{$\delta t$}                          & $\beta =$             & $0.52 \pm 0.04$        & $0.55 \pm 0.03$        & $0.52 \pm 0.12$        & $0.47 \pm 0.02$         & $0.55 \pm 0.04$        \\ \hline
\multirow{2}{*}{\textbf{$V_{\Delta t}^{\rm SOL}$}} & $\beta =$             & $0.44 \pm 0.04$        &             & $0.32 \pm 0.05$        & $0.34 \pm 0.06$         &             \\ \cline{2-7} 
                                      & $\gamma =$         &             & $2.24 \pm 0.1$       &             &              & $2.10 \pm 0.15$       \\ \hline
\multirow{2}{*}{\textbf{$V_{\Delta t}^{\rm USD}$}} & $\beta =$            & $0.36 \pm 0.06$        &             & $0.3 \pm 0.03$        & $0.31 \pm 0.03$         &             \\ \cline{2-7} 
                                      & $\gamma =$         & $2 \pm 0.13$    & $1.94 \pm 0.11$       &             &              & $2.1 \pm 0.15$    \\ \hline
\end{tabular}
\end{table}

\subsection{Temporal correlations}

\subsubsection{Autocorrelation function}

\begin{figure}
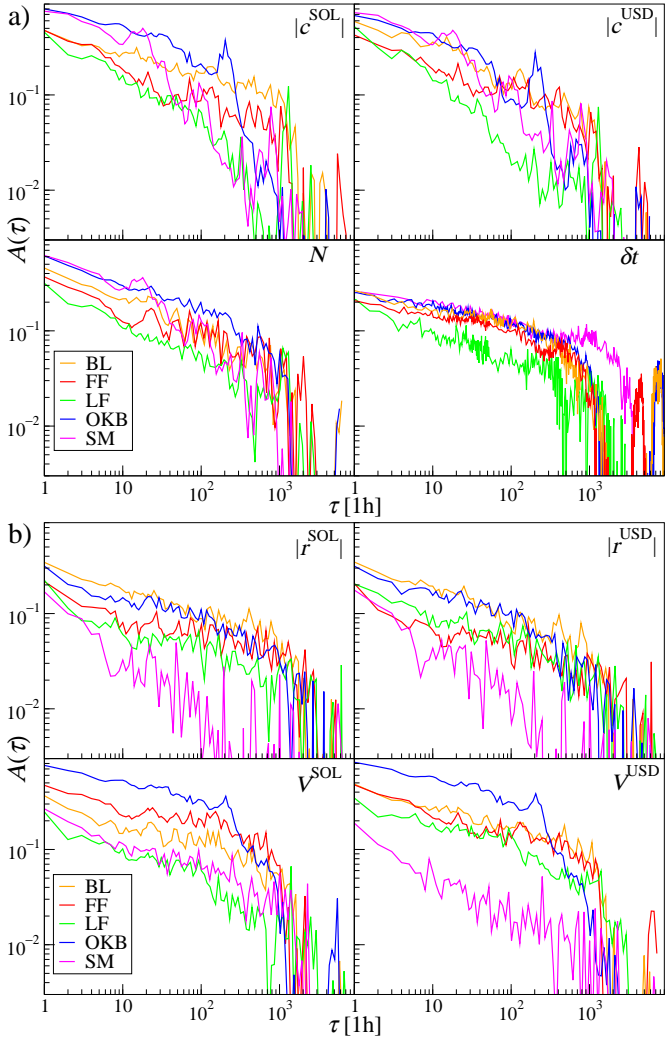

\includegraphics[width=0.49\textwidth]{figs/ACF_kolekcjeuu.eps}

\includegraphics[width=0.49\textwidth]{figs/ACF_floor_price.eps}
\caption{(a) Autocorrelation function for absolute values of logarithmic increments of collection capitalization expressed in solana $|c_{\Delta t}^{^{\rm SOL}}|$ (top left) and US dollar $|c_{\Delta t}^{^{\rm USD}}|$ (top right), the number of transactions aggregated hourly $N_{\Delta t}$ (bottom left), and inter-transaction times $\delta t$ (bottom right). (b) Autocorrelation function for absolute values of floor price returns expressed in SOL $|r_{\Delta t}^{^{\rm SOL}}|$ (top left) and USD $|r_{\Delta t}^{^{\rm USD}}|$ (top right) and volume value expressed in solana $V_{\Delta t}^{^{\rm SOL}}$ (bottom left) and US dollar $V_{\Delta t}^{^{\rm USD}}$ (bottom right).}
\label{fig::acf}
\end{figure}

Temporal relationships in time series can involve both linear and non-linear dependencies. The former are quantified in terms of autocorrelation function
\begin{equation}
A(x,\Delta i) = { 1/T \sum_{i=1}^T \left[ x(i) - \langle x(i) \rangle_i \right] \left[ x(i+\Delta i) - \langle x(i) \rangle_i \right] \over \sigma^2_x},
\end{equation}
where $\sigma_x^2$ denotes variance of a time series ${x(i)}$, $\langle \cdot \rangle$ denotes mean, and $\Delta i$ denotes temporal lag that is measured in points of the time series (its clock-time equivalent is $\tau = \Delta i \Delta t$). Fig.~\ref{fig::acf} shows the autocorrelation functions calculated for each time series and for each of the analysed collections. In all cases, there is a slow decay of $A(x,\tau)$, which reflects the existence of long-range autocorrelations. For some time series, it is clear from the graphs that we observe a power-law decay. The lifetime of these long-range autocorrelations is observable-dependent -- the longest ones are observed for inter-transaction time $\delta t$ and for the floor price absolute returns $|r_{\Delta t}|$, reaching approximately 1,000 h (i.e., more than a month). If we now look at Fig.~\ref{fig::stat}, we see that an average cluster lasts for a comparable amount of time, which makes both observations mutually consistent (see Fig.~6 in Ref.~\cite{DrozdzS-2009a}). The strength of autocorrelation varies between the time series and it can be pointed out that the observables for some collections show systematically stronger autocorrelation than for the other ones: these include OK Bears and Blocksmith Lab Smyths.

\subsubsection{Hurst exponent}

Long-term memory that manifests itself in a power-law decay of the autocorrelation function introduces non-stationarity to a time series. On the other hand, autocorrelation function is substantially sensitive to data nonstationarity, which means that one needs another measure of temporal correlations which can be used in such a case. A tool to quantify correlations in the nonstationary data is the Hurst exponent $H$; it can be calculated by means of the detrended fluctuation analysis (i.e., MFDFA with $q=2$, see Sect.~\ref{sect::data.and.methods}). The Hurst exponent allows for distinguishing between the long-term positively correlated (persistent) time series ($0.5 < H \leqslant 1$), the long-term negatively correlated (antipersistent) time series ($0 \leqslant H < 0.5$), and uncorrelated ones ($H=0.5$). By using the formalism of MFDFA, we calculated $H$ for each time series of interest by fitting a power-law model to the respective fluctuation function $F_q(s)$ for $q=2$ (see the distinguished green lines in Figs.~\ref{fig::Fq.1} and~\ref{fig::Fq.2}). The corresponding numerical results are collected in Tab.~\ref{tab::hurst}.

Majority of the observables for all the collections show a persistent behaviour with $0.6 \leqslant H \leqslant 0.8$ -- this is true for the capitalization increments, the number of trades, inter-transaction times, and transaction volume value. For the unsigned observables like $\delta t$, $N_{\Delta t}$, and $V_{\Delta t}$, it is an expected, natural result if one takes into consideration the positive autocorrelations reported above. For the signed observables like $c_{\Delta t}$, however, this outcome is less expected. It can be accounted for by the fact that, even if separated by a substantial lag, the transaction prices of different tokens from the same collection go in the same direction, which leads to a certain inertia in the evolution of its capitalization. This inertia causes, in turn, the observed persistence. In contrast, a slight yet statistically significant antipersistence ($0.42 \leqslant H \leqslant 0.47$) can be detected for the floor price returns in a few cases: $r_{\Delta t}^{\rm SOL}$ for BL, FF, LF, and SM and $r_{\Delta t}^{USD}$ for FF and SM. The origin of this effect is low liquidity and long inter-transaction times that cause most non-zero returns to be isolated in a time series. It resembles the analogous results for bitcoin in its early years~\cite{DrozdzS-2018a,watorek2021} and for regular assets like stocks and currencies at very short time scales $\Delta t$~\cite{DrozdzS-2010a}. The remaining cases of the floor price return time series seem to be uncorrelated to within statistical error -- see Tab.~\ref{tab::hurst}. The latter result matches the analogous results reported for the price returns in the other financial markets~\cite{KwapienJ-2012a}, which supports a view that the floor price for NFTs can be considered as the closest possible counterpart of the transaction price in the case of other assets.

\begin{table}[]
\caption{Values of the Hurst exponent together with its standard error calculated for time series representing each observable and collection.}
\label{tab::Hurstexp}
\scriptsize
\begin{tabular}{|l|c|c|c|c|c|}
\hline
     & BL & FF & LF & OKB & SM \\ \hline
\textbf{$c_{\Delta t}^{\rm SOL}$} & $0.67\pm0.02$      & $0.84 \pm 0.04$         & $0.61  \pm 0.03$       & $0.77 \pm 0.04$          & $0.81 \pm 0.04$         \\ \hline
\textbf{$c_{\Delta t}^{\rm USD}$} & $0.68 \pm 0.03$         & $0.82 \pm 0.02$    & $0.57\pm 0.05$         & $0.79 \pm 0.04$          & $0.81 \pm 0.04$         \\ \hline
\textbf{$r_{\Delta t}^{\rm SOL}$} & $0,47 \pm 0.01$  & $0.47 \pm 0.01$   & $0.45 \pm 0.01$          & $0.49 \pm 0.01$   & $0.42 \pm 0.01$          \\ \hline
\textbf{$r_{\Delta t}^{\rm USD}$} & $0.49 \pm 0.01$   & $0.47 \pm 0.01$   & $ 0.49 \pm 0.01$   & $0.51 \pm 0.01$    & $0.46 \pm 0.01$       \\ \hline
\textbf{$N_{\Delta t}$}       & $0.74 \pm 0.02$  & $0.69 \pm 0.01$   & $0.62 \pm 0.01$     & $0.81 \pm 0.01$   & $0.76 \pm 0.02$         \\ \hline
\textbf{$\delta t$}       & $0.65 \pm 0.01$    & $0.66 \pm 0.01$  & $0.68 \pm 0.01$  & $0.70 \pm 0.01$  & $0.68 \pm 0.01$    \\ \hline
\textbf{$V_{\Delta t}^{\rm SOL}$} & $0.65 \pm 0.01$    & $0.74 \pm 0.02$   & $0.65 \pm 0.02$ & $0.82 \pm 0.03$  & $0.67 \pm 0.01$ \\ \hline
\textbf{$V_{\Delta t}^{\rm USD}$} & $0.69 \pm 0.01$   & $0.8 \pm 0.01$  & $0.69 \pm 0.03$  & $0.82\pm 0.04$  & $0.66 \pm 0.02$   \\ \hline
\end{tabular}
\label{tab::hurst}
\end{table}

\subsection{Multifractal properties}
The existence of nonlinear long-range correlations (Fig.~\ref{fig::acf}) indicates a possibility of multifractal structures. We thus applied MFDFA to the time series of logarithmic increments of capitalization, floor price returns, transaction volume value aggregated hourly, the number of transactions aggregated hourly, and inter-transaction times for each collection. In any multifractal approach, the fundamental issue is identification of scaling in a plot of the relevant function. In MFDFA it is the fluctuation function given by Eq.~(\ref{eq::fluctuation.function}). We calculated $F_q(s)$ for all 8 types of data for each collection.

\begin{figure}
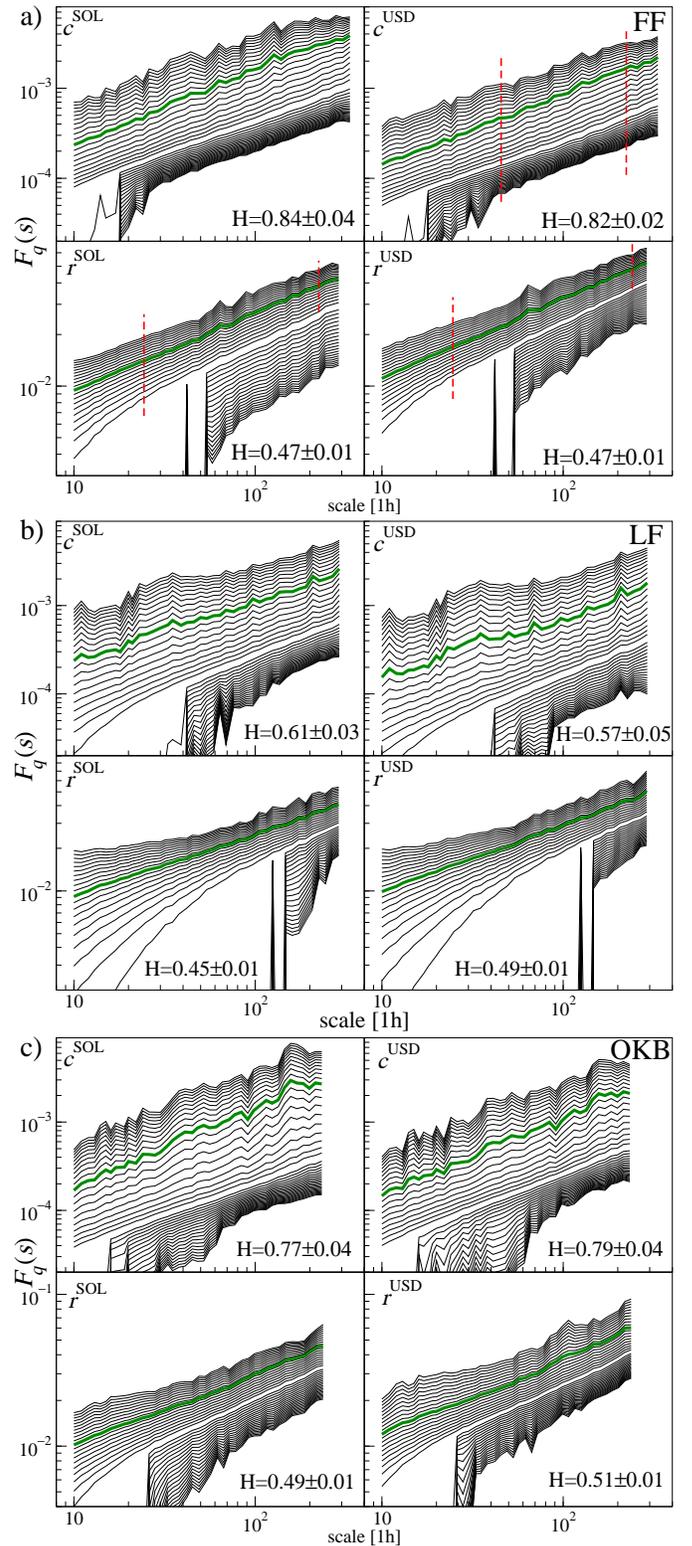

\includegraphics[width=0.49\textwidth]{figs/Fq_FF_index_floorpriceplusuH.eps}
\includegraphics[width=0.49\textwidth]{figs/Fq_LF_index_floorpriceplusuH.eps}
\includegraphics[width=0.49\textwidth]{figs/Fq_OKB_index_floorpriceplusuH.eps}
\caption{Multifractal analysis of time series representing three NFT collections: (a) FF, (b) LF, and (c) OKB. The respective fluctuation functions $F_q(s)$ with $-4 \leqslant q \leqslant 4$ were calculated for the logarithmic increments of collection capitalization expressed in solana $c_{\Delta t}^{^{\rm SOL}}$ and US dollar $c_{\Delta t}^{^{\rm USD}}$ and floor price returns expressed in the same currencies: $r_{\Delta t}^{^{\rm SOL}}$ and $r_{\Delta t}^{^{\rm USD}}$. A range of scales, in which a power-law form of $F_q(s)$ is observed for a range of values of $q$, is denoted by vertical red dashed lines on each plot. The heavy green lines in each panel correspond to $F_q(s)$ with $q=2$ and serve calculation of the Hurst exponents $H$ (the range-of-scales restrictions do not apply here).}
\label{fig::Fq.1}
\end{figure}

Let us start from capitalization and floor price return time series. Fig.~\ref{fig::Fq.1} presents the fluctuation function plots for 3 selected collections: FF, LF, and OKB, which constitute the most characteristic cases. It comes straightforward that there is no time series that would reveal a firm power-law dependence over at least a decade of scales and for all the considered values of $q$ simultaneously. There is only one example of a time series that shows scaling over both positive and negative $q$s: $c_{\Delta t}^{^{\rm USD}}$ for the FF collection; even its counterpart for SOL fails to do so as regards $q > 0$. However, this time series does not develop a scaling range that would be at least a decade long, which means that fractal structure is present only to a rather limited extent there. It happens, though, that scaling can be seen over a broader scale range but only for $q > 0$; these are $r_{\Delta t}^{^{\rm SOL}}$ and $r_{\Delta t}^{^{\rm USD}}$ for the FF collection. As the positive $q$s allow the algorithm to select medium and large fluctuations, we conclude that they reveal fractality, indeed, while small fluctuations contain only non-stationary noise (because stationary noise could be fractal as well). Such a conclusion is not surprising at all as in many financial markets small orders and the related small variations of observables are associated with noise traders that do not contribute genuinely to the overall dynamics. Apart from this, there are cases where a scaling range is too short or not observed at all, so no existence of a uniform fractal structure may be inferred: $c_{\Delta t}^{^{\rm SOL}}$ for all the three collections and $c_{\Delta t}^{^{\rm USD}}$ for LF and OKB, as well as $r_{\Delta t}^{^{\rm SOL}}$  for LF and OKB and $r_{\Delta t}^{^{\rm USD}}$ for LF and OKB.

We would like to stress here that our results obtained with MFDFA are to be considered as indicative. The main outcome is that, although the NFT market is young and the least developed of all financial markets, the NFT data considered in this work show a remarkably rich internal structure that may even be multifractal. Due to several factors, the most important of which is the limited amount of data available as individual time series~\cite{DrozdzS-2009a,KwapienJ-2023a}, we state it here as a likely possibility, while we shall leave its quantitative verification to future works.

\begin{figure}
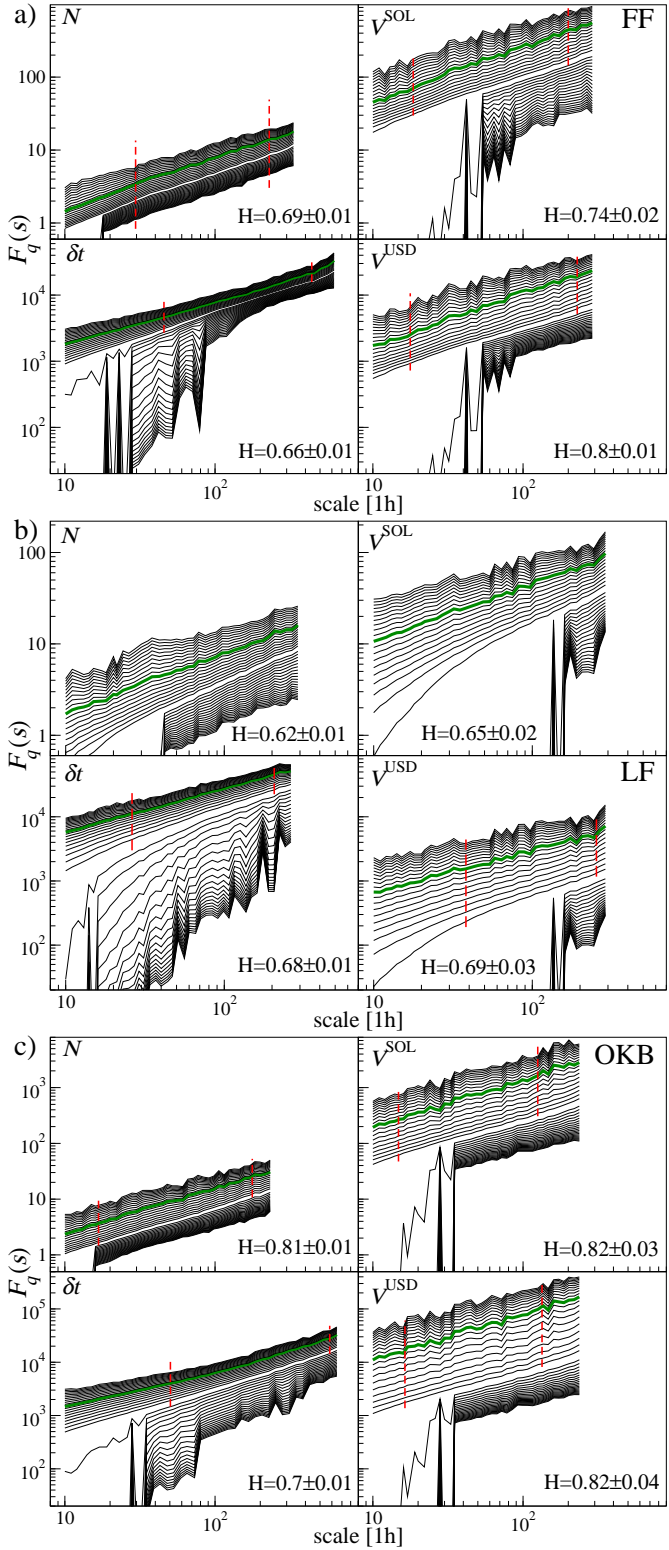

\includegraphics[width=0.49\textwidth]{figs/Fq_FF_V_N_TdplusH.eps}
\includegraphics[width=0.49\textwidth]{figs/Fq_LF_V_N_TdplusuH.eps}
\includegraphics[width=0.49\textwidth]{figs/Fq_OKB_V_N_TdplusuH.eps}
\caption{The same as in Fig.~\ref{fig::Fq.1} but for different quantities: the number of transactions aggregated hourly $N_{\Delta t}$, inter-transaction times $\delta t$, and transaction volume value aggregated hourly expressed in solana $V_{\Delta t}^{^{\rm SOL}}$ and US dollar $V_{\Delta t}^{^{\rm USD}}$. The heavy green lines in each panel correspond to $F_q(s)$ with $q=2$ and serve calculation of the Hurst exponents $H$ (the range-of-scales restrictions do not apply here).}
\label{fig::Fq.2}
\end{figure}

The fluctuation functions for the remaining 4 observables: the number of transactions $N_{\Delta t}$, inter-transaction times, and transaction volume value in SOL and USD are collected in Fig.~\ref{fig::Fq.2}. There are two time series that show scaling over both positive and negative $q$s: $N_{\Delta t}$ for FF and OKB. The time series for which a range of power-law scaling is minimum acceptable are: $V_{\Delta t}^{^{\rm SOL}}$ for FF and OKB as well as $V_{\Delta t}^{^{\rm USD}}$ and $\delta t$ for all the three collections. The remaining time series develop either too short scaling range or their fluctuation functions do not scale at all.

\begin{figure}
\includegraphics[width=0.49\textwidth]{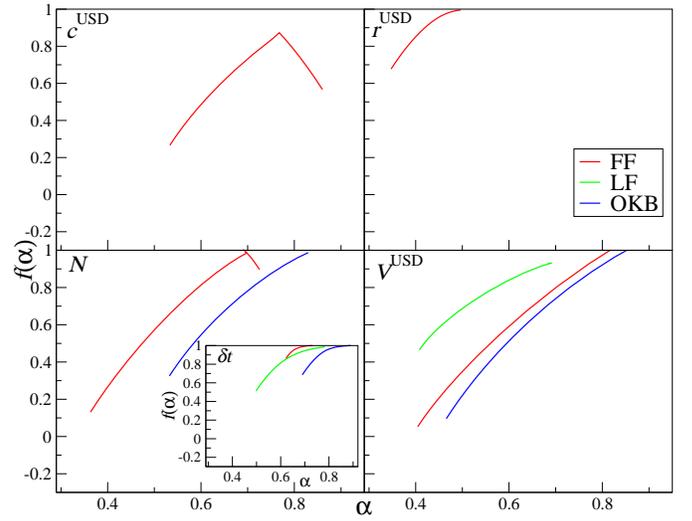}
\caption{Singularity spectra $f(\alpha)$ calculated for time series that reveal scaling in their fluctuation functions $F_q(s)$. In most cases this corresponds to using $q > 0$.}
\label{fig::spectra}
\end{figure}

The nature of the power-law behaviour of $F_q(s)$ for different values of $q$ indicate that the corresponding time series carry some signatures of multifractality. This can be inferred directly from Fig.~\ref{fig::spectra} where singularity spectra $f(\alpha)$ are shown for all the cases in which $F_q(s)$ reveals scaling over at least a decade in Fig.~\ref{fig::spectra}. The resulting spectra are typically left-sided which indicates multifractal structure observed for medium and large values of the corresponding time series. Only in two cases, $N_{\Delta t}$ and $c_{\Delta t}^{^{\rm USD}}$ for FF, a right arm of $f(\alpha)$ reflecting the organization of small values of the corresponding observable can be determined. However, this arm is shorter than the left one, which makes the spectra left-side asymmetric. It has to be noted in this context that multifractal organization of certain observables like price returns, inter-transaction times, and transaction volume values is a well-established result in financial data and constitutes one of its stylized facts~\cite{ContR-2001a,ChakrabortiA-2011a,SCHINCKUS2014,Kutner2019}. The present results obtained for the NFT collections show a less developed multifractality, origin of which we shall attribute to worse market liquidity and poorer data statistics. Somewhat different principles of the NFT market dynamics may also play some role in this connection. The related attributes may however change when this market gains more trader interest in the future.

The plots shown in Figs.~\ref{fig::Fq.1}-\ref{fig::Fq.2} contain two features, which are repetitive across the time series and which ha
values of $F_q(s)$ for small scales if $q < 0$. It comes from the fact that the considered time series change their values less frequently than the assumed sampling frequency $\Delta t=1$ h, which produces zero-value periods that significantly lower values of the fluctuation function for $q < 0$ and throw them out of the plots. This can be avoided by appropriately adjusting the minimum scale considered in an analysis if possible. However, in the present case the required cut-offs would be too large for the plots to be even created, so we decided not to do it. The second feature to be pointed out is a broom-like effect for the short scales seen the most clearly for the floor price returns in the LF case (Fig.~\ref{fig::Fq.1}(b)). Such a shape of the fluctuation functions is typical for time series containing outliers if there is no multifractality~\cite{ChenZ-2022a,DrozdzS-2009a,Oswiecimka2020,watorek2021,KwapienJ-2023a}.

\section{Conclusions}

Unlike the relatively well-understood cryptocurrency market, the NFT market is still in its early stage of development and does not attract as much research interest. Our work is one of the first to analyze the statistical properties of selected observables characterizing instruments traded on the NFT market. The data consisted of tick-by-tick data and floor price quotes for 5 selected token collections created on the Solana network: Blocksmith Labs Smyths, Famous Fox Federation, Lifinity Flares, Okay Bears, and Solana Monkey Business. From this data, we created for each collection a set of time series representing 7 observables recorded every hour each: logarithmic increments of the total collection capitalization, floor price returns, transaction volume values, the number of transactions, and inter-transaction times. The values, which are expressed natively in SOL units, were also recalculated and expressed in USD to relate their value to fiat money.

The time series we created were examined in terms of probability density distributions, a possible occurrence of autocorrelation, values of the Hurst exponent, and a potential fractal structure. Probability density distributions turned out to have heavy tails for all the types of time series considered. The behaviour of these tails was modeled with a power function and a stretched exponential function. In the case of the latter, the compliance of the empirical data with the model turned out to be satisfactory for the inter-transaction times, the number of transactions in time unit, and the floor price returns expressed in both USD and SOL (all the collections), the transaction volume value expressed in both USD and SOL (LF and OKB), the USD-expressed capitalization increments (OKB), and the SOL-expressed ones (BL and OKB). For the USD- and SOL-expressed capitalization increments, a model that was able to reproduce the data for LF relatively well was the power-law model with $\gamma \approx -3/2$, while for the USD- and SOL-expressed transaction volume values, the same model with $\gamma \approx 2$ was appropriate for FF and SM. It sometimes happens, however, that both models can reproduce the empirical data with comparable accuracy. Such a result cannot be considered as unexpected, because curvature of a stretched exponential function can be remarkably broad and go unnoticed if only an insufficiently long part of this function is shown on a double logarithmic plot. This was the case for the capitalization increments expressed in USD (BL and FF) and in SOL (FF) as well as for the transaction volume value expressed in USD (BL). In the remaining cases, on the other hand, neither of these two models was optimal.

These outcomes have to be compared with the corresponding features of the regular financial markets and the cryptocurrency market. On those markets the heavy-tail pdfs are a common property for many quantities like price returns or volume: the tails are the heavier, the shorter is the time scale considered. At the shortest time scales the inverse cubic scaling for price returns is observed which is counted among the financial stylized facts. By taking into consideration how new is the NFT market as compared to the other financial markets and even to the cryptocurrency one, it is noteworthy that the power laws can be observed there for certain data sets. On the other hand, the stretched exponential distribution fits well to the distribution of inter-transaction times, which also occurs in the case of cryptocurrencies.

All the tested time series are characterized by long-range correlations, reaching up to two months. This is a much longer period than in the case of other financial instruments, such as stocks or fiat currencies. Such a long correlation duration may be related to the low liquidity of the NFT market. In individual cases, the identified correlations are close to power-law. It is worth stressing that, although the Pearson autocorrelation function was used, the observed autocorrelations must be viewed as nonlinear ones wherever the modulus of the original time series was considered (the floor price returns and capitalization increments). The Hurst exponents revealed substantial persistence in the capitalization increments, the number of transactions in time unit, the inter-transaction times, and the transaction volume values, which supports the results obtained with the autocorrelation function. Only the floor price returns were slightly antipersistent for BL, FF, and SM, while close to being independent for LF and OKB. This outcome resembles parallel results reported in literature for bitcoin price returns in the early phase of bitcoin trading. The analysis carried out with MFDFA allowed us for the identification of a fractal structure in the case of the time series of logarithmic increments of capitalization expressed in USD and floor price returns (in both SOL and USD) for FF, the number of transactions for FF and OKB, inter-transaction times and transaction volume values expressed in USD for all the three collections, and the transaction volume values expressed in SOL for FF and OKB. For these fractal time series, the quality of scaling was satisfactory enough that we could identify a trace of multifractality (which is directly related to nonlinear correlations). In all the remaining cases no fractal structure could be found at all. One cannot decide whether this result has to be attributed to the finite-size effects that substantially affect the fluctuation functions calculated with MFDFA or to genuine property of the NFT market dynamics. It requires further studies to resolve this issue.

Our analysis has shown that, in some respects, the NFT market somewhat differs from other financial markets, including the cryptocurrency one. In particular, we found neither the inverse cubic scaling of the returns nor consistent, uniform multiscaling of the considered observables. It must be noted in this context, however, that our analysis included only a single sampling frequency of 1 hour, therefore we cannot exclude a possibility that the inverse cubic behavior may be found for some other time scale. We also cannot exclude a possibility that multiscaling could be identified if the studied time series were substantially longer, which would improve statistics. Based on our results, we can conclude that the NFT market is still in its early phase. Regardless of this conclusion, it is also important to stress the limited number of NFT collections examined. It is thus advisable to expand the data set, both on other collections operated by the Solana network and on collections operated by other networks (e.g. Ethereum). This indicates directions of the related future work.

\section*{Data Availability Statement}

The data are freely available from CryptoSlam! portal (tick-by-tick transaction data)~\cite{CryptoSlam} and Magic Eden portal (floor price data)~\cite{MagicEden}.

\nocite{*}
\bibliography{ref}

\end{document}